\documentclass{revtex4}

\usepackage{graphicx}
\newcommand{\msbar}{$\overline{\rm MS}$}
\newcommand{\lsim}{\buildrel < \over {_\sim}}

\begin{document}
\bibliographystyle{revtex}

\preprint{UPR--965--T}
\preprint{E3012}

\title{Determinations of $\alpha(M_Z)$: Comparison and Prospects}



\author{Jens Erler}
\email[]{erler@ginger.hep.upenn.edu}
\affiliation{Department of Physics and Astronomy, University of Pennsylvania,
             Philadelphia, PA 19146-6396, USA}


\date{\today}

\begin{abstract}
I review and compare various techniques to obtain the value of the QED 
coupling, $\alpha$, at the $Z$ pole.  GigaZ precisions would require
a much more accurate determination than available today.  A combination
of the virtues of current methods may help to achieve this goal.
\end{abstract}

\maketitle


The value of the QED coupling constant at $Z$ pole energies,
\begin{equation}
\alpha (M_Z) = {\alpha\over 1 - \Delta\alpha(M_Z)},
\end{equation}
continues to induce the dominant theoretical uncertainty in the interpretation
of the observables from LEP~1 and the SLC. A future linear $e^+ e^-$ collider with 
GigaZ~\cite{Hawkings:1999ac,Erler:2000jg} capability would be able to greatly 
improve the current measurements, with $\alpha (M_Z)$ ultimately dominating 
the overall uncertainty.  Therefore, it would be essential to improve the 
present error, $\delta\alpha (M_Z)/\alpha(M_Z) \approx \pm 2 \times 10^{-4}$,
by at least a factor of two or three.  While the discussion will be limited
to $\alpha (M_Z)$, in its essence it also applies to the two-loop hadronic 
uncertainty in the muon anomalous magnetic moment. 

All methods and renormalization schemes to determine $\alpha (M_Z)$ utilize 
experimental data up to some cut-off, $s_{\rm cut}$, beyond which perturbative 
QCD (PQCD) is evoked. They differ due to differences in the data sets and 
treatments; the choice of $s_{\rm cut}$; the reference renormalization scheme; 
the option to add experimental information from 
$\tau$-decays~\cite{Alemany:1998tn} (assuming isospin invariance); space-like 
vs.\ time-like integration; the treatment of heavy quarks; the use of a QCD 
sum rule~\cite{Groote:1998pk} and/or a resummation optimization; and so on.

In terms of the photon polarization function, $\Pi_\gamma (s)$, 
$\Delta\alpha$ is given by,
\begin{equation}
\Delta\alpha(s) = - 4\pi\alpha 
  \left[ \Pi_\gamma^\prime(s) - \Pi_\gamma^\prime(0) \right] 
  \hspace{10pt} \mbox{(on-shell scheme),} \hspace{30pt}
  \Delta\hat\alpha(\mu) = 4\pi\alpha \hat\Pi_\gamma (0) 
  \hspace{10pt} \mbox{(\msbar\ scheme)}.
\end{equation}
At the one-loop level in perturbation theory one finds for a fermion of charge
$Q_f$ ($N_c^f$ is the color factor),
\begin{equation}
  \Delta\alpha = {\alpha\over 3\pi} \sum_f Q_f^2 N_c^f \left[ \ln {s\over m_f^2}
  - {5\over 3} \right] \hspace{10pt} \mbox{(on-shell),} \hspace{30pt}
  \Delta\hat\alpha(\mu) = {\alpha\over 3\pi} \sum_f Q_f^2 N_c^f 
  \ln {\mu^2\over m_f^2} \hspace{10pt} \mbox{(\msbar)}.
\end{equation}
Numerically, $\alpha^{-1} (M_Z) \sim 129$ and $\hat\alpha^{-1} (M_Z) \sim 128$.
Alternatively, the on-shell quantity can be represented by a once subtracted
dispersion relation (SDR),
\begin{equation}
  \Delta\alpha (s) = - 4\alpha s \int\limits_{4 m_\pi^2}^\infty d s^\prime 
 {{\rm Im}\Pi_\gamma^\prime(s^\prime)\over s^\prime(s^\prime - s - i\epsilon)}.
\label{DI}
\end{equation}
In the case of hadrons, $R(s) = 12\pi {\rm Im} \Pi_\gamma^\prime (s)$, and in 
the standard SDR approach one has,
\begin{equation}
  \Delta\alpha^{(5)}_{\rm had} (M_Z^2) = - {\alpha M_Z^2\over 3\pi} \left[
  \underbrace{\int\limits_{4 m_\pi^2}^{s_{\rm cut}}
  {R(s) d s\over s(s - M_Z^2 - i \epsilon)}}_{\rm DATA} +
  \underbrace{\int\limits_{s_{\rm cut}}^\infty
  {R(s) d s\over s(s - M_Z^2 - i \epsilon)}}_{\rm PQCD} \right],
\label{SDR}
\end{equation}
where the superscript indicates application to all quarks except the top.
In the \msbar\ scheme it is more natural to work with an unsubtracted
dispersion relation (UDR)~\cite{Erler:1999sy},
\begin{equation}
  \Delta\hat\alpha^{(3)}_{\rm had} (s_{\rm cut}) = {\alpha\over 3\pi}
  \underbrace{\int\limits_{4 m_\pi^2}^{s_{\rm cut}} 
  {R(s) d s\over s - i \epsilon}}_{\rm DATA} + 2 \alpha 
  \underbrace{\int\limits_0^{2\pi} d \theta\, 
  \hat\Pi_\gamma^{(3)}(\theta)}_{\rm PQCD},
\label{UDR}
\end{equation}
where the second integral is along a circle with radius $s = s_{\rm cut}$.
Since typical values for $\sqrt{s_{\rm cut}}$ are 1.8~GeV~\cite{Davier:1998kw}
and 2.5~GeV~\cite{{Eidelman:1999vc}}, one only needs to include the three 
light quarks in Eq.~(\ref{UDR}). One then uses an analytical 
solution~\cite{Erler:1999sy} to the order $\alpha\alpha_s^3$ and 
$\alpha^2$ renormalization group evolution (RGE) to {\em decrease\/} 
$\mu^2 = s_{\rm cut}$ to $\mu^2 = \hat{m}_c^2 (\hat{m}_c)$ (the \msbar\ charm mass) 
where one matches the effective field theories with three and four effective quark flavors.
The matching is performed at order $\alpha\alpha_s^2$ at which subtle effects 
from internal charm quark loops have to be taken into account. However, well 
below the charmonium threshold these are small and strongly decoupling despite 
$\hat{m}_c < s_{\rm cut}$. Then one evolves up in energy and includes 
the $\tau$-lepton and the $b$-quark. However, this is successful {\em only\/} 
when a short-distance quark mass definition (such as \msbar) is used.  
Transition to the on-shell mass definition would introduce large $\pi^2$ terms,
rendering application to bottom (charm) quarks questionable (impossible). 
Thus, in the UDR approach bottom and charm effects can be described 
{\em entirely\/} within PQCD, avoiding complications at heavy quark resonances. 
On the other hand, in the SDR approach one has to abandon PQCD in the vicinity
of resonance regions. 

Focussing on only one quark flavor at a time, one could relate 
the integral expression of the SDR approach to the analytical 
expressions~\cite{Erler:1999sy} of the UDR approach.  The resulting equation
has the form of a specific type of QCD sum rule which could be used to
{\em determine\/} $\hat{m}_c$ and $\hat{m}_b$. However, this is only
the first entry in an infinite series of sum rules~\cite{Jamin:1997rt} --- and 
not the one which uses the data most efficiently.  This implies that combining
the UDR approach with an appropriate QCD sum rule is a recipe to minimize 
the uncertainties from the $b$ and $c$ quark sector~\cite{Erler02}. 
\begin{table}[t]
\caption{Comparison of QCD analyses.  The values and uncertainties quoted in
the original papers are adjusted to $\alpha_s (M_Z) = 0.120$ (fixed).
The quark mass uncertainty in the BF-MOM scheme is from the pole masses
which cannot be improved.  The UDR approach uses \msbar\ masses; their 
error can be expected to decrease significantly in the future. 
The theory errors in the SDR and UDR approaches include the uncertainty 
introduced by assuming quark-hadron duality near the cut-off of the dispersion
integrals.  The PQCD error in the BF-MOM approach has not been estimated, yet.}
\label{comp}
\vspace{10pt}
\begin{tabular}{l|c|c|c}
                                      & SDR     & BF-MOM  & UDR          \\
\hline
quantity                              & $R(s)$  & $D(-s)$ & $\beta(\mu)$ \\
$\Delta\alpha_{\rm had}^{(5)}$        & 0.02770 & 0.02773 & 0.02779      \\
quoted uncertainty                    & 0.00015 & 0.00018 & 0.00020      \\
$\alpha_s$-dependence                 & linear approximation & not available & fully analytic \\
contribution from J/$\Psi$ resonances & 65\%    & 15\%    & 0\%          \\
error from quark masses               & 0       & 0.00010 & 0.00015      \\
error from data                       & 0.00015 & 0.00015 & 0.00011      \\
theory error                          & 0.00002 & (0)     & 0.00007      \\ 
$s_{\rm cut}$                         & 1.8 GeV & 2.5 GeV & 1.8 GeV      \\
reference                             & \cite{Davier:1998si} & \cite{Jegerlehner:2001ca} & \cite{Erler:1999sy} \\
\end{tabular}
\end{table}
Another advantage of the UDR approach is that all theoretical contributions are
available as explicit analytical expressions, with no need for a numerical
integration.  In particular, the $\alpha_s$ and quark mass dependences are all
taken into account.  This is important for global analyses in which these
parameters enter in many different places, causing non-trivial and non-linear
correlations.  In the SDR approach only a crude linear 
approximation~\cite{Davier:1998si} is available.

Another way to reduce the impact of the resonance region is to use the analytic
structure of $\Pi_\gamma$ and to work in the Euclidean (space-like) 
region~\cite{Eidelman:1999vc},
\begin{equation}
  \Delta\alpha^{(5)}(M_Z^2) = \underbrace{\left[ \Delta\alpha^{(5)}(M_Z^2) - 
  \Delta\alpha^{(5)}(-M_Z^2) \right]}_{\rm PQCD} + \underbrace{\left[ 
  \Delta\alpha^{(5)}(-M_Z^2) - 
  \Delta\alpha^{(5)}(-s_{\rm cut}) \right]}_{\rm PQCD} +
  \underbrace{\Delta\alpha^{(5)}(-s_{\rm cut})}_{\rm DATA}.
\label{BFMOM}
\end{equation}
The first term is the analytical continuation from the Minkowski (time-like) 
to the Euklidean region. The second term represents the RGE in the perturbative
Euklidean domain in which $R(s)$ is replaced by the Adler $D$ function.
It is computed in the {\em gauge dependent\/} background 
field momentum subtraction (BF-MOM) scheme up to three-loop 
order~\cite{Jegerlehner:1999zg}. Unlike the \msbar\ scheme, the BF-MOM scheme 
is a mass-dependent renormalization scheme. Thus both, the UDR and BF-MOM 
approaches, depend explicitly on the quark masses. In the latter, the quark pole
masses are used.  This is disadvantageous since a long-distance mass definition
such as the pole mass has an intrinsic renormalon ambiguity of order 
$\Lambda_{\rm QCD}$. The heavy quark sector also contributes via the last term 
in Eq.~(\ref{BFMOM}) where, like in the SDR approach, the resonance region 
complicates the analysis. However, in the BF-MOM approach the resonance 
contribution is suppressed by about a factor of four~\cite{Jegerlehner:2001ca}.
Note, that in this approach there may be a subtle correlation between the 
uncertainty from the theory (quark masses) and data (resonance region) parts. 

The advantage of splitting the data and theory parts as in Eq.~(\ref{BFMOM}) is
that no reference to global or local quark hadron duality is needed. 
In contrast, the SDR and UDR approaches both have an explicit momentum cut-off 
where the transition from data (hadrons) to theory (quarks) occurs. 
In principle, this could give rise to a significant cut-off dependence, 
especially when non-perturbative (NP) effects produce a strongly oscillating 
form of the hadronic cross section, $i.e.\ R(s)$. Such oscillations 
arise neither in PQCD nor in the operator product expansion (OPE) which 
accounts for a certain class of NP effects. Clearly, the cut-off dependence 
should be kept small. However, it is not necessarily optimal to demand that it
vanishes. Indeed, the importance of duality violating effects has been overemphasized 
in the past, and there are good reasons to believe that they are 
small~\cite{Bigi:2001ys}. While unidentified sources of (OPE breaking) NP 
effects are hazardous, they are not the only source of uncertainty. For 
example, a poorly converging perturbative expansion can be even more perilous.

Table~\ref{comp} summarizes the comparison of the three methods defined by
Eqs.~(\ref{SDR}), (\ref{UDR}), and (\ref{BFMOM}).  Table~\ref{theory}
gives a breakdown of the theory error in $\alpha^{-1} (M_Z)$ in the UDR approach.
The corresponding error in $\Delta\alpha$ is obtained by multiplying by $\alpha$.

The available data in the region $m_c \lsim \sqrt{s} \lsim 2 m_c$ are rather 
poor, and one would like to replace them by a robust theoretical description as
far as possible. It is therefore desirable to improve the UDR approach by 
lowering the cut-off dependence and to decrease ones exposure to duality 
violating effects.  Conversely, the BF-MOM approach would benefit by utilizing 
a short-distance mass definition.  Furthermore, as a matter of practice, it may
be unwieldy to properly correlate the uncertainties from resonances and quark 
masses; this may eventually result in a somewhat larger uncertainty compared
to the UDR approach.  Nevertheless, it appears that within both, the UDR and 
BF-MOM methods, one has the potential to obtain a solid theory driven evaluation
of $\alpha (M_Z)$. In contrast, the more traditional SDR approach with its
strong reliance on the complicated function $R(s)$ seems inadequate for 
the high demands of GigaZ precisions. 

\begin{table}[t]
\caption{Error breakdown in the UDR approach. The quoted non-perturbative QCD 
uncertainty is due to OPE breaking effects typically of the form $e^{-C/\alpha_s}$, 
where $C$ is in general complex leading to an oscillating $R(s)$.  
Davier and H\"ocker~\cite{Davier:1998si} fit a variety of oscillating curves 
to the experimental $R(s)$ around $s_{\rm cut}$ and conclude 
$\Delta \alpha^{-1} = \pm 0.002$. Here I use a more conservative 
estimate~\cite{Erler:1999sy}. There are other non-perturbative (higher twist)
effects {\em within\/} the OPE. These are of
${\cal O} (\alpha_s^2/\pi^2 \Lambda_{\rm QCD}^4/s^2_{\rm cut}; 
\alpha_s^2/\pi^2 m_K^2 f_\pi^2/s^2_{\rm cut}) \sim 2\times 10^{-7}$ and of
${\cal O} (\Lambda_{\rm QCD}^4/m_c^4; \Lambda_{\rm QCD}^4/m_b^4) \sim -3 \dots -7 \times 10^{-6}$, 
respectively, and can be neglected.  The parametric error due to the imperfect knowledge
or $\alpha_s$ is excluded here; the $\alpha_s$ dependence is fully included in electroweak fits.}
\label{theory}
\begin{tabular}{r|c|c|c}
& sector & uncertainty & comment \\
\hline
perturbative QCD                      & $u$, $d$, $s$ & 0.005 & missing ${\cal O}(\alpha\alpha_s^3)$ corrections \\
perturbative QCD                      & $c$, $b$      & 0.004 & missing ${\cal O}(\alpha\alpha_s^3)$ corrections \\
perturbative QCD                      & RGE           & 0.003 & missing ${\cal O}(\alpha\alpha_s^4)$ corrections \\
non-perturbative QCD                  & $u$, $d$, $s$ & 0.006 & quark-hadron duality \\
\hline
total QCD                             & $u$, $d$, $s$, $c$, $b$ & 0.009 & theory \\
\hline
\msbar\ quark mass                    & $c$ & 0.019 & $\hat{m}_c = 1.31 \pm 0.07$~GeV \\
\msbar\ quark mass                    & $b$ & 0.002 & $\hat{m}_b = 4.24 \pm 0.11$~GeV \\
\hline
total quark masses                    & $c$, $b$      & 0.021 & parametric \\
\hline
$R(s)$ \cite{Davier:1998si}           & $u$, $d$, $s$ & 0.015 & data \\
\hline
grand total                           & $u$, $d$, $s$, $c$, $b$ & 0.027 & data + theory + parametric \\
\end{tabular}
\end{table}


\end{document}